\documentclass[12pt]{article}
\usepackage{graphicx}
\small

\begin{document}

\centerline{\bf KODAIKANAL CALCIUM IMAGES: DETECTION OF PLAGES, FIXING}

\vskip 0.3cm
\centerline{\bf THE HELIOGRAPHIC COORDINATES AND ESTIMATION OF AREA}

\vskip 0.3cm
\centerline {\em K. M. Hiremath$^{1,2}$, Shreyam Krishna$^{3}$, Adithya H. N$^{4}$, S. R. Chinmaya$^{5}$ and Shashanka R Gurumath$^{6}$}
\vskip 0.3cm
1. Formerly Indian Institute of Astrophysics, Bangalore, India
Emai: hiremath@iiap.res.in

2. \#23, Mathru Pithru Krupa, 2nd Cross, 1st Main, BDA Layout, Bikasipura, BSK V Stage, Bengaluru-560111, India

3. B-9, Sector H, Aliganj Near Sangam Crossing Lucknow, Uttar Pradesh, India

4. \#76, Neelanduru Nalluru, Sringeri, Chikkamagaluru, Karnataka, India 

5. \#3, First floor, 12th Main road, Lakkasandra extension.  Bangalore 560030, India

6. Physical Research Laboratory, Navrangpura, Ahmedabad - 380009, India

\vskip 0.3cm

\centerline{ \bf ABSTRACT}

Kodaikanal Observatory is a veritable treasure trove
of data, with the data repository covering almost 100 years of observations. For the years 1909-2007, we use
 calibrated Ca II K spectroheliograms from the Kodaikanal Observatory to
 detect the plages, fix their heliographic coordinates and also estimate the plage
areas. We adopt the following procedure. After ensuring that, for all the
 years,  Kodai calcium images have very negligible ellipcity,
a circle is fitted and two central coordinates and radius of calcium images
are determined uniquely. For each pixel of the calcium image,
we then fix heliographic coordinates and  extract plages along with their weighted average
coordinates.
The heliographic coordinates of these extracted plages are then compared with the heliographic coordinates of photospheric sunspots
from the Greenwich sunspot database and chromospheric magnetic plages detected from the SOHO/MDI magnetograms.
We find that the heliographic
coordinates of calcium plages match very well with the heliographic coordinates of sunspots and magnetic plages
authenticating our method of detection of plages and computation of positional coordinates.
 A code is developed in Python and all the nearly century scale plages data,
with accurately estimated heliographic coordinates and areas, is available to the public.

\section{INTRODUCTION}

Two unsolved mysteries of the sun are genesis of solar cycle and activity phenomena
and how sun's irradiance varies with the short and long term time scales. Although
 so called flux transport dynamo models apparently reproduce butterfly
diagram, question remains how to reproduce the integrated solar cycle and activity
phenomena such as sunspots, coronal holes, faculae, plages, {\em etc}. Other important and pertinent question
is whether solar cycle and activity phenomena can be understood from the flux
transport dynamo models or superposition of long period ($\sim$ 11 yrs or more)
 MHD waves that probably might have generated and might have traveled along
the fossil magnetic field structure (Alfv{\'e}n 1943, Hiremath 1994, Hiremath 1995).
Infact, from the century scale sunspot data of the Greenwich photoheliographic
results, Gokhale {\em et.al.} (1990)
came to the conclusion that sunspots might have originated from the superposition
of long period MHD waves. Hiremath (2010) showed that well known solar periods
can be reproduced from the perturbation of combined poloidal and toroidal magnetic
field structure embedded in the sun's interior, probably of primordial origin.

For validation of both the turbulent dynamo
and MHD oscillation models and to test which model is most consistent and appropriate
to explain the combined solar cycle and activity phenomena, observational information
regarding sun's long term ( $\ge$ 10 yrs) magnetic activity is required. Although
magnetic activity information inferred from the magnetograms is nearly five decades
old, sun's activity on century scale that may yield clues regarding mystery of
Maunder minimum type of activity warrants nearly century scale or greater than
time scale of magnetogram data.

Recent overwhelming evidences (Hiremath 2009, Hiremath 2015) are building up that influence of sun's activity
phenomena on the Earth's climate can not be neglected so easily. If we understand
this sun-Earth climate relationship, we can also understand the stars-planets climate
relationship and ultimately to search for habitable exoplanets in the universe.
Hence, for deep understanding of sun-Earth's climate relationship, information
regarding long-term variation of solar irradiance is required.

Considering the importance of long-term variations of sun's magnetic  and irradiance
activities, Kodaikanal calcium  data is ideally suited. In Kodaikanal, we have a treasure
trove of both the white light and calcium image data observed from the same relevant instruments
with very good seeing conditions. As there is one to one correspondence between
(Ortiz and Rast 2005, Sivaraman and Livingston 1982, Rodono {\em et.al.} 1987, Frasca {\em et.al.} 2000, Butler 1995, Frasca {\em et.al.} 2008, 
Loukitcheva {\em et.al.} 2009, Frasca {\em et.al.} 2010, Bertello {\em et.al.} 2016a, Bertello {\em et,al.} 2016b, 
Pevtsov {\em et.al.} 2016) calcium plages and
photospheric magnetic activity, calcium plages can be
used as proxy for understanding the long term variation of magnetic activity
of the sun. As chromospheric activity is one of the main contributor for the solar
irradiance, analysis and extraction of chromospheric activity parameters are
 very useful for modeling and reconstruction of long-term solar irradiance variations.

Aims of present study are to: i) uniquely estimate the center and radius of the
calcium image, ii) reasonably detect boundary of the plages, iii) from {\em ab initio}, accurately
fix the heliographic coordinates for all the pixels of the image and compute
the average heliographic coordinates (such as latitude and longitude from the central meridian) of plages and, iv) as presented
in the following sections, validate position (heliographic coordinates) of Kodai calcium plages
with the position of photospheric magnetic sunspots and chromospheric magnetic
plages and, v) validate the estimated projectional corrected calcium plage areas
with the chromospheric magnetic plage areas.

From the extracted heliographic coordinates, in future, we plan to investigate the nature
of latitudinal and long-term ($\sim$ century scale) temporal variation of chromospheric rotation rate profile. Cycle to
cycle variation and long-term  changes in the velocity of meridional
circulation (for which accurate estimation
of heliographic coordinates is required) and its relationship with the photospheric
and deep interior meridional circulation will also be investigated.

Present study is not a first study to delineate the plages
from the calcium images, fix the heliographic coordinates and
estimate the areas. However, most of the previous studies (Foukal 1996, Balmaceda {\em et.al.} 2009,
Ermolli {\em et.al.} 2009a, Ermolli {\em et.al.} 2009b, Tlatov {\em et.al.} 2009, Bertello {\em et.al.} 2010, Priyal {\em et.al.} 2014, Chatterjee {\em et.al} 2016, Chatzistergos 2017, Priyal {\em et.al.} 2017, Chatzistergos {\em et.al.} 2018, Barata {\em et.al.} 2018)
 mainly concentrated on  estimating hemispheric contribution of plage areas.
Although some of these studies attempted to estimate  the heliographic
coordinates of the plages, clarity regarding method of estimation of heliographic coordinates
and their accuracy is lacking. In addition, most of these studies do not apply the
projectional corrections for the estimated areas of the plages. Present study
fulfill both of these objectives.

Plan of the paper is as follows. In section 2, data and analysis of the Kodai calcium image
data are presented. In section 3, the results are presented and,
last section consists of the conclusions emerged from this study.
                                                                             
\section{DATA AND ANALYSIS}

The Kodaikanal Solar Observatory has been observing the Sun in the Ca II K line since
1905 by using photographic plates illuminated by 30 cm objective telescope.
With a spatial resolution of about 2 arc sec, calcium
narrow band filter is used to obtain the spectroheliograms in Ca II K line. This data repository
spans over  hundred years, from 1905 - 2007. Recently nearly century scale data of calcium
images is digitized by 4096 x 4096 pixels, with a pixel size of 0.85 arc seconds.
Details of the telescope, digitization and calibration can be found
in the previous (Priyal {\em et.al.} 2014, Priyal {\em et.al.} 2017) studies.
 Typical calibrated calcium image is illustrated in Figure 1a.

\begin{figure}
\begin{center}
     Figure 1a \hskip 30ex  Figure 1b
%\vskip -6.5ex
    \begin{tabular}{cc}
      {\includegraphics[width=13pc,height=12pc]{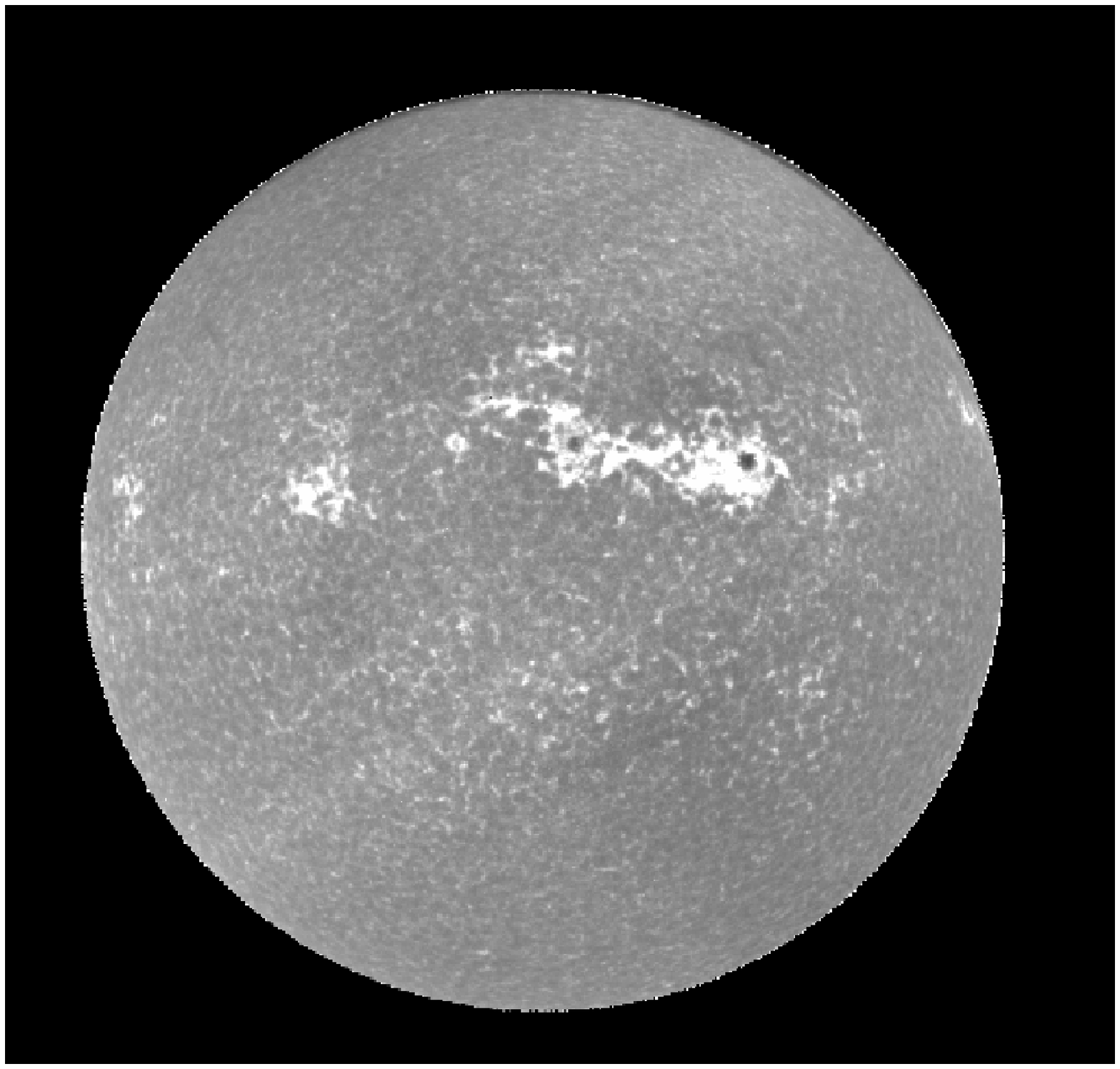}} &
      %\vector(10,10){15}
      {\includegraphics[width=13pc,height=12pc]{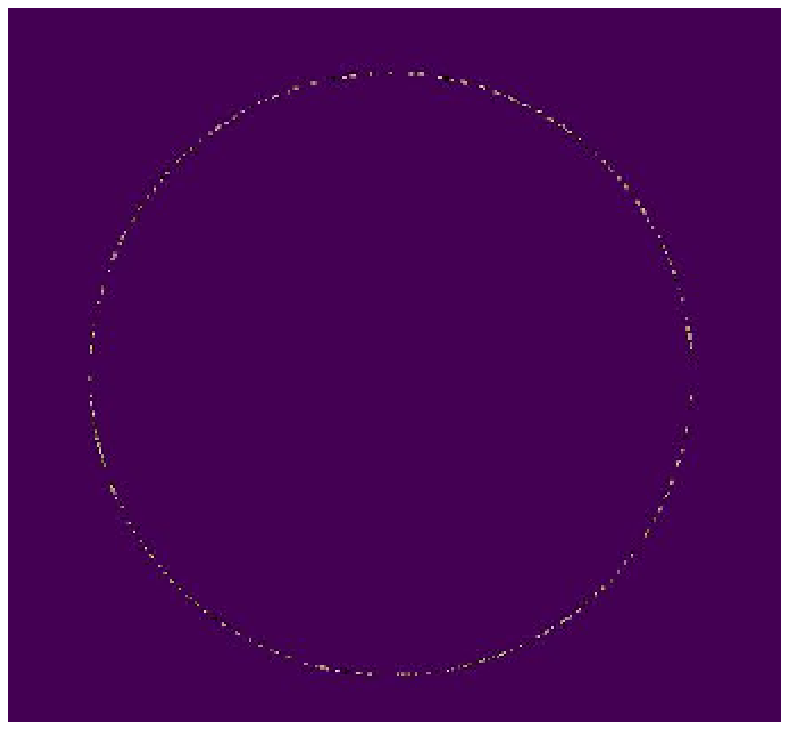}} \\
\end{tabular}
    \caption{ Figure 1a illustrates the calibrated Ca II K Spectroheliogram from the Kodaikanal
Observatory data. Where as Figure 1b illustrates the detected edge of the solar image.}
\end{center}
\end{figure}

\subsection{ Methods Of Analysis Of Solar Images}

\subsubsection {Edge Detection}

 The method we have used to find the radius and the center is solving three simultaneous
equations that have three unknowns (two central coordinates and
one radial coordinate). Hence by solving three equations with three unknowns,
the center and the radius of circle are estimated uniquely. We consider
every data point in the detected solar disk edge for least-square fitting.
The added advantage is, we have taken all the pixels of the detected
edge and circle is fitted for the sun, which is better estimation than
taking a few data points as used in most of the previous studies.

The edge of the sun is detected using OpenCV’s canny edge detection function which consists of
4 parts:

\noindent a) 5x5 Gaussian filter is used for the removal of the noise in the images.

\noindent b) Sobel filter is used to get vertical and horizontal gradients from smoothed image.
 From these two horizontal and vertical gradient images, edge gradient $G$ and angle $\theta$ for each pixel
are obtained by using the following formulae:

\begin{equation}
G = ({G_{x}}^{2} + {G_{y}}^2)^{1/2} ,
\end{equation}
\noindent and
\begin{equation}
 \theta = tan^{-1} ({G_{y}\over G_{x}}) .
\end{equation}
\noindent where $G_{x}$ is the gradient in x direction and $G_{y}$ is the gradient in y direction
which are perpendicular to edges.

\noindent c)  In order to remove any unrelated pixels that are not part of the edge, each pixel is checked in the direction of the gradient to confirm that it is the local maximum in its neighborhood.

\noindent d) First minimum and maximum intensity values of edges are estimated.
Then any edge intensity greater than maximum intensity values are considered
to be sure edges, otherwise are considered to be of non-edges. Whereas the edge intensities
that lie between maximum and minimum intensity edge values are classified as either
edges or no-edges depending on their connectivity. Connected edges are considered
to be ``sure-edge" pixels which are subset of edges, otherwise are discarded.
This step is known as Hysteresis Thresholding.

Canny edge detection was chosen as the preferred method after comparing it with Sobel x,
Sobel y and Laplacian edge detection methods. The comparison revealed that the error in
detection of edge was lowest in canny edge detection. Following all these steps, detected
edge of the calcium image is illustrated in Figure 1b.

\subsubsection{ Circle Fitting}

All the detected pixels of edge of the image is least-square fitted with a circle that uses a system of
simultaneous three equations with three unknowns to get a unique solution for center and radius of the solar image. This
method only requires that the edge of the circle coordinates be the input and initial guess is
not required. This method computes uniquely the three required coordinates (two coordinates for
center of the circle and the radius). The process is described in detail below:

If $x_{i}$ and $y_{i}$ (where $i=1,N$, $N$ is total number) are Cartesian coordinates
of the detected pixels and, $\bar{x}$ and $\bar{y}$  are their
respective means which are defined as follows

\begin{equation}
\bar{x} = {\sum_{i} x_{i} \over {N}} ,
\end{equation}
and
\begin{equation}
\bar{y} = {\sum_{i} y_{i} \over {N}} .
\end{equation}
Let $x_{i}$ and $y_{i}$ be further transferred into new variables $u_{i}$, $v_{i}$ such that
\begin{equation}
u_{i} = x_{i} - \bar{x} ,
\end{equation}
and
\begin{equation}
v_{i} = y_{i} - \bar{y} .
\end{equation}
Let ($u_{c}, v_{c}$) be the center coordinates of the circle with radius $R$ and $\alpha = R^{2}$.

\noindent Distance of any point ($u_{i}, v_{i}$) from the center is $= [(u_{i}-u_{c})^{2}+ (v_{i} -  v_{c})^{2}]^{1/2}$.

\noindent From the method of least-square fit which implies function $S = \sum_{i} [g(u_{i} , v_{i} )]^{2}$
should be minimum wherein
\begin{equation}
g(u_{i},v_{i}) = (u_{i} - u_{c})^{2} + (v_{i} - v_{c})^{2} - \alpha. 
\end{equation}

\noindent Hence,  partial derivatives of this
function with respect to $\alpha$, $u_{c}$ and $v_{c}$ should all be zero. For the partial derivative
of $S$ with respect to $\alpha$ we get

\begin{equation}
{\partial S \over {\partial \alpha}} = 2 \sum_{i} {g(u_{i},v_{i}) {\partial g \over {\partial \alpha}}}= 0 , 
\end{equation}

\begin{equation}
\Rightarrow -2 × \sum_{i} g (u_{i} , v_{i} ) = 0 ,
\end{equation}

\begin{equation}
 \Rightarrow  \sum_{i} [(u_{i} - u_{c})^{2} + (v_{i} - v_{c})^{2} - \alpha] = 0 ,
\end{equation}

\begin{equation}
\Rightarrow  \sum_{i} {u_{i}}^{2} + \sum_{i} {v_{i}}^{2}+ \sum_{i} {u_{c}}^{2} + \sum_{i} {v_{c}}^{2} - 2[\sum_{i} u_{i} u_{c}
 + \sum_{i} v_{i} v_{c} ] = \sum_{i} \alpha ,
\end{equation}

\begin{equation}
\Rightarrow [\sum_{i} {u_{i}}^{2} + \sum_{i} {v_{i}}^{2}]+ N [{u_{c}}^{2} + {v_{c}}^{2} ] - 2[u_{c} \sum_{i} u_{i} + v_{c} 
\sum_{i} v_{i} ] = N \alpha .
\end{equation}

\noindent With a known fact that $\sum_{i} u_{i} = \sum_{i}(x_{i} - \bar{x}) = N \bar{x} - N \bar{x} = 0$  and, $\sum_{i} v_{i} = 0$,
we get

\begin{equation}
\Rightarrow \sum_{i} {u_{i}}^{2} + \sum_{i} {v_{i}}^{2}+ N [{u_{c}}^{2} + {v_{c}}^{2}] = N \alpha . 
\end{equation}

\noindent For the partial derivative of $S$ with respect to $u_{c}$ we get

\begin{equation}
{{\partial S} \over {\partial u_{c}}} = 2 × \sum_{i} {g (u_{i} , v_{i} ) {\partial g \over {\partial u_{c}}}} = 0 ,
\end{equation}

\begin{equation}
\Rightarrow \sum_{i}(u_{i} - u_{c} )g(u_{i} , v_{i} ) = 0 .
\end{equation}

\noindent On expansion, we get

\begin{eqnarray}
\Rightarrow \sum_{i} {u_{i}}^{3} + \sum_{i} u_{i} {v_{i}}^{2} - 2u_{c} \sum_{i} {u_{i}}^{2} - 2v_{c} \sum_{i}
u_{i} v_{i} - u_{c} \sum_{i} {u_{i}}^{ 2} \\
- u_{c} \sum_{i} {v_{i}}^{ 2} - N {u_{c}}^{ 3} - N u_{c} {v_{c}}^{ 2} + N \alpha u_{c} &=& 0
\end{eqnarray}

\noindent By substituting the value $N \alpha$ from equation (13), we get
\begin{equation}
u_{c} \sum_{i} {u_{i}}^{ 2} + v_{c} \sum_{i} u_{i} v_{i} = {1\over{2}} [\sum_{i} {u_{i}}^{ 3} + \sum_{i} u_{i} {v_{i}}^{ 2} ] 
\end{equation}

\noindent Lastly partial derivative of $S$ with respect to $v_{c}$ yields the following equation

\begin{equation}
{{\partial S} \over {\partial v_{c}}} = 2 × \sum_{i} {g [u_{i} , v_{i} ] {\partial g \over {\partial v_{c}}}} = 0 ,
\end{equation}

\noindent Following the derivations of above equations (8)-(13), we obtain the following equation:
\begin{equation}
u_{c} \sum_{i} u_{i} v_{i} + v_{c} \sum_{i} {v_{i}}^{ 2} = {1\over{2}} [ \sum_{i} {v_{i}}^{ 3} + \sum_{i} v_{i} {u_{i}}^{ 2}  ] .
\end{equation}

\noindent Simultaneous solution of equations (18) and (20) yield the two central coordinates $u_{c}$ and $v_{c}$. Then
from equation (13), we get the value of $\alpha$ ($=R^{2}$) and, hence the radius R.

\noindent By adding two central coordinates $(u_{c},v_{c})$ to the respective means, original central coordinates $x_{c}$ and $y_{c}$
are obtained

\begin{equation}
x_{c} = u_{c} + \bar{x} ,
\end{equation}
and
\begin{equation}
y_{c} = v_{c} + \bar{y}.
\end{equation}
\noindent Hence, importance of this method is from the observed sun's image, two central coordinates
and its radius are obtained uniquely.

\begin{figure}
\begin{center}
     Figure 2a \hskip 30ex  Figure 2b
%\vskip -6.5ex
    \begin{tabular}{cc}
      {\includegraphics[width=12pc,height=12pc]{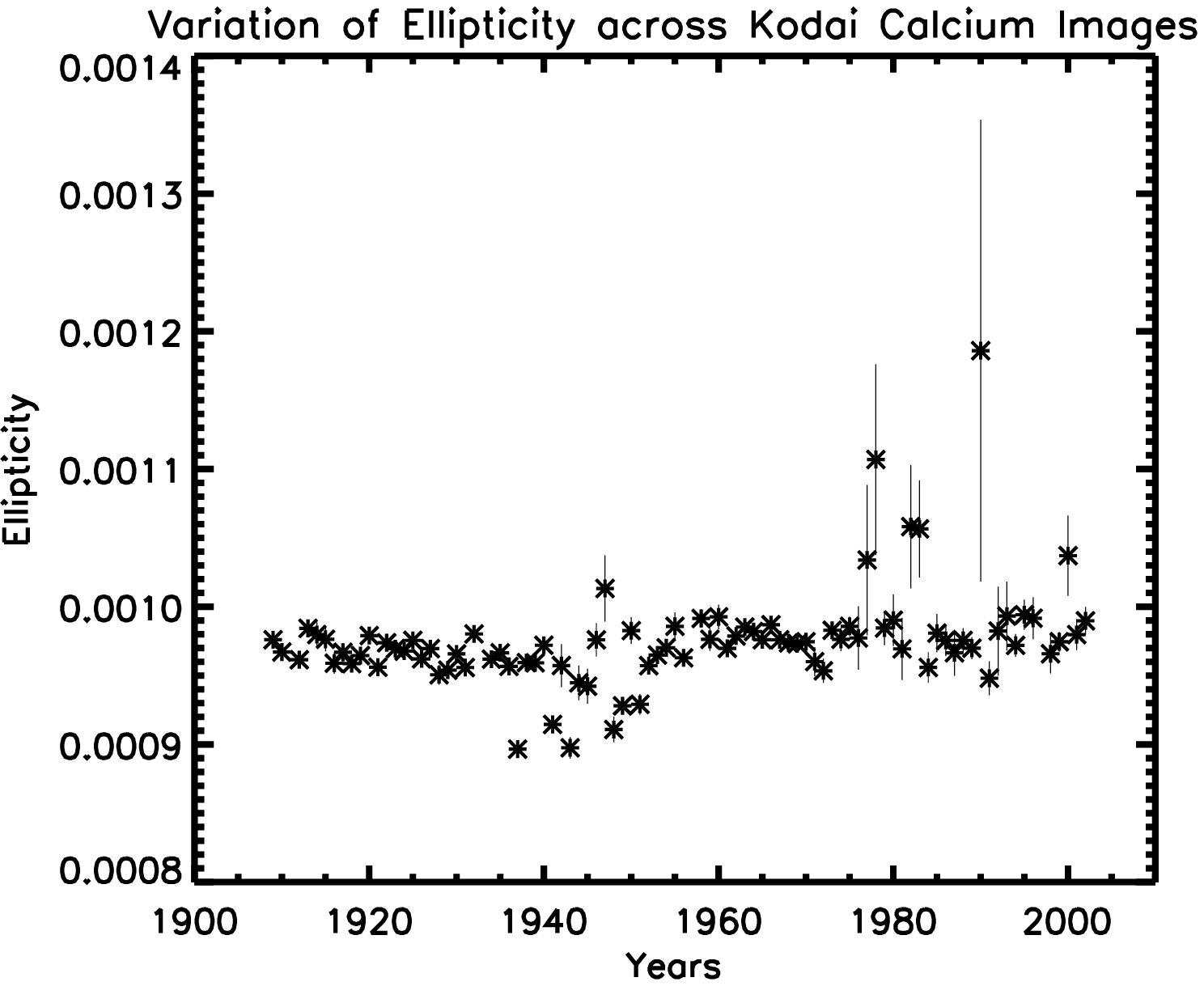}} &
      %\vector(10,10){15}
      {\includegraphics[width=13pc,height=12pc]{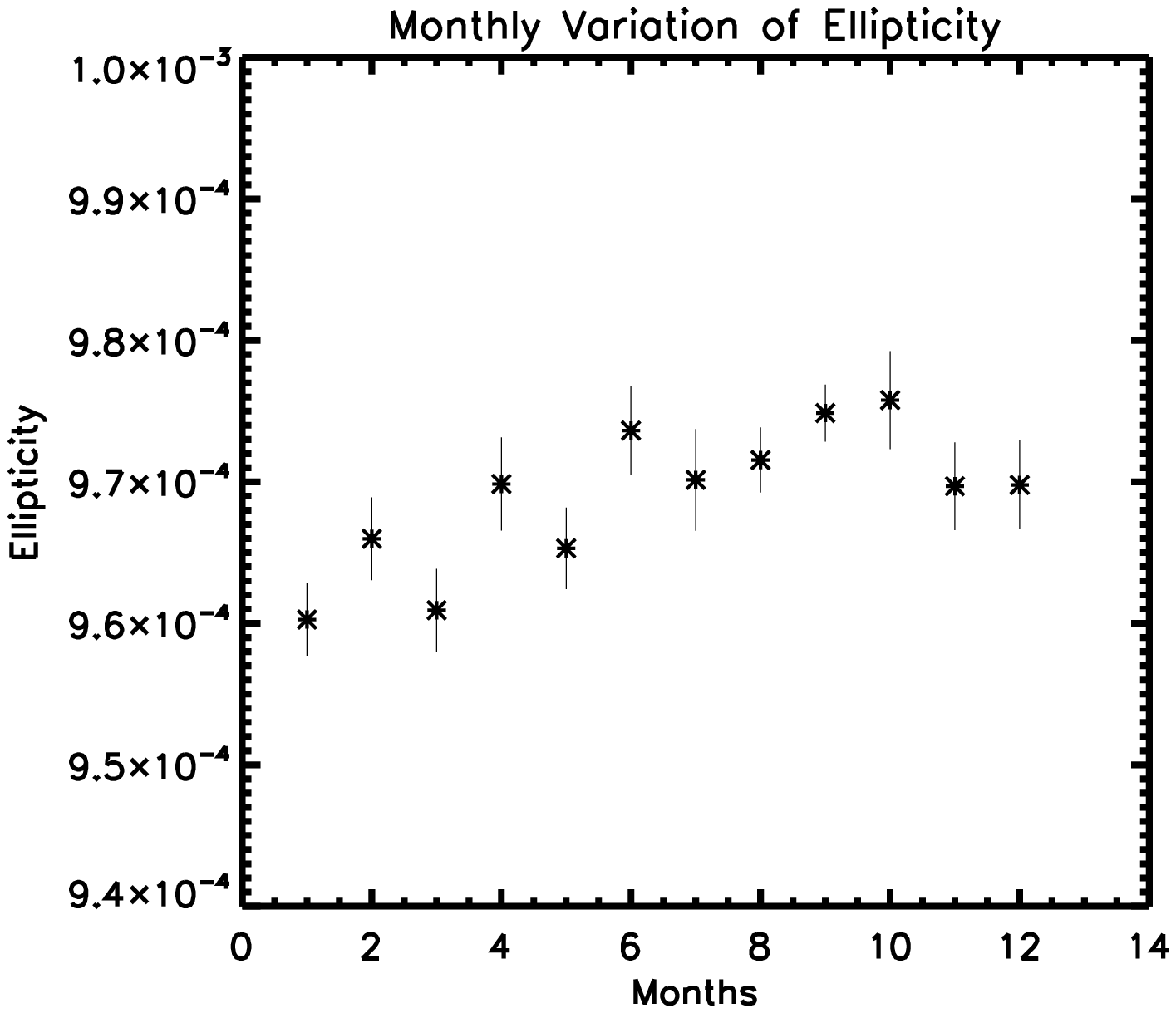}} \\
\end{tabular}
    \caption{ Figure 2a illustrates estimated annual ellipcity and whereas Figure 2b illustrates
the same parameter for different months of a year.}
\end{center}
\end{figure}

\begin{figure}
\begin{center}
     Figure 3a \hskip 30ex  Figure 3b
%\vskip -6.5ex
    \begin{tabular}{cc}
      {\includegraphics[width=12pc,height=12pc]{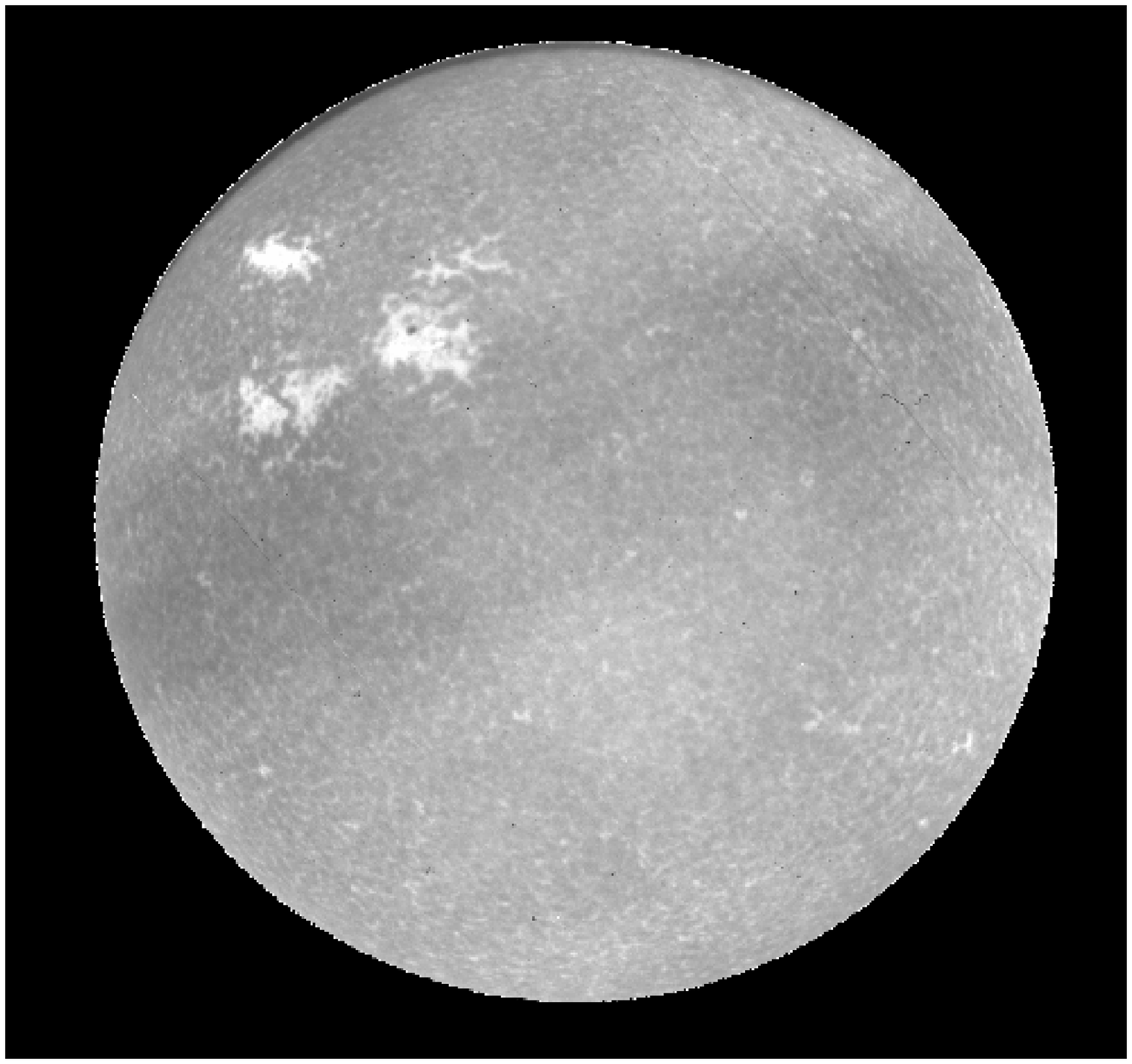}} &
      %\vector(10,10){15}
      {\includegraphics[width=12pc,height=12pc]{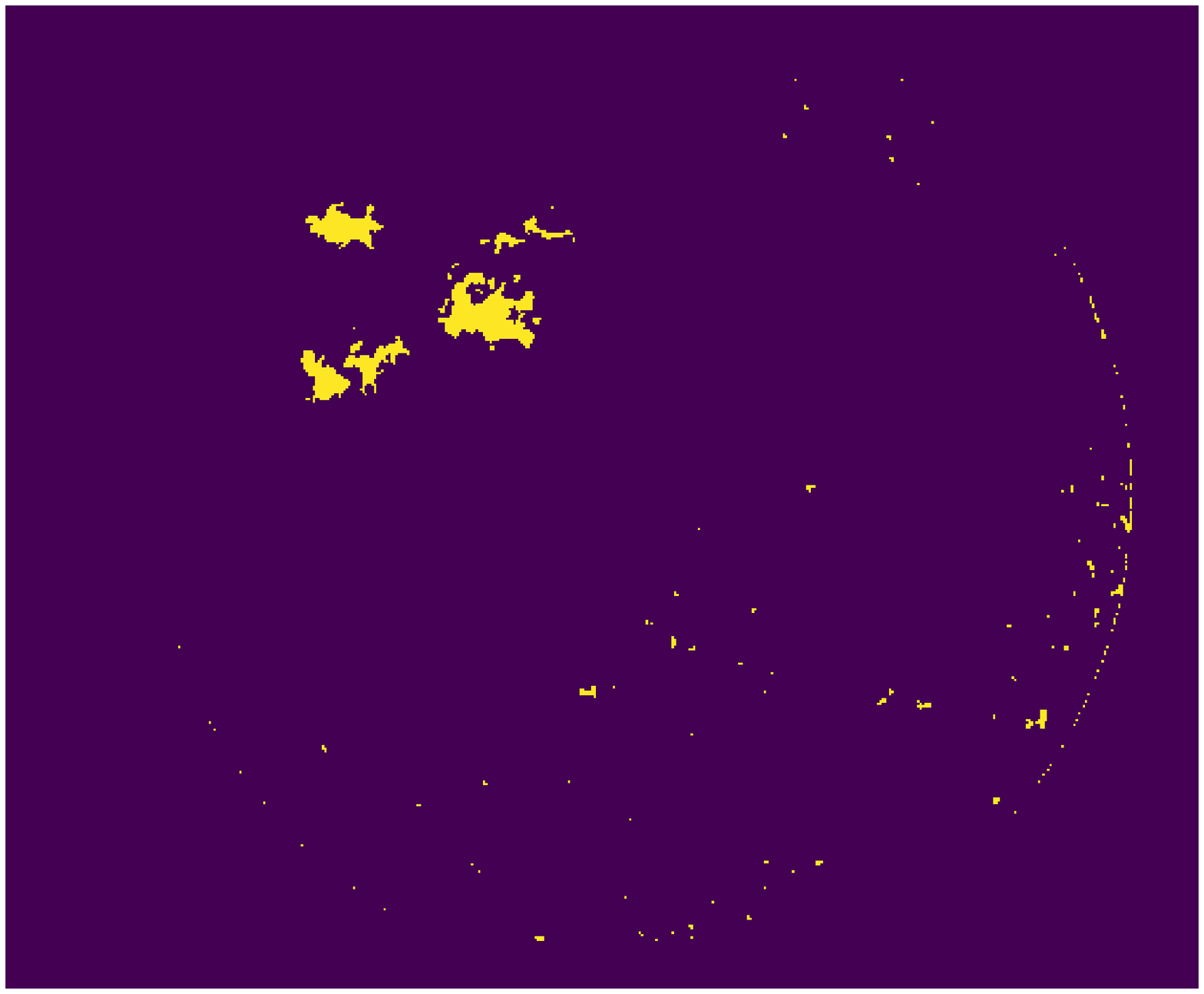}} \\
\end{tabular}
    \caption{ Figure 3a illustrates the original calcium image,  whereas, Figure 3b illustrates
the image with detected plages.}
\end{center}
\end{figure}

\begin{figure}
\begin{center}
      Figure 4a \hskip 30ex  Figure 4b
%\vskip -6.5ex
    \begin{tabular}{cc}
      {\includegraphics[width=12pc,height=11pc]{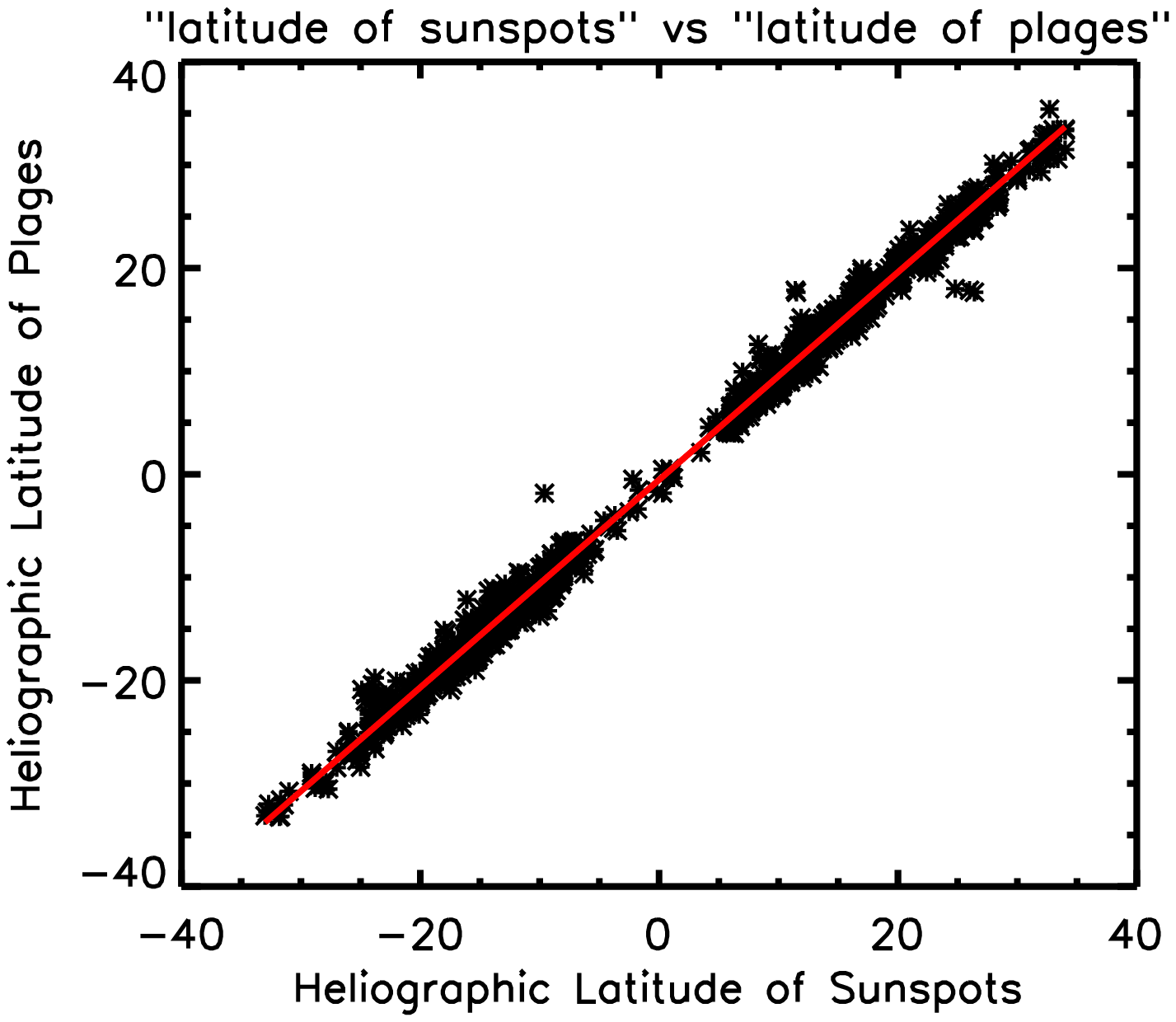}} &
      %\vector(10,10){15}
      {\includegraphics[width=13pc,height=11pc]{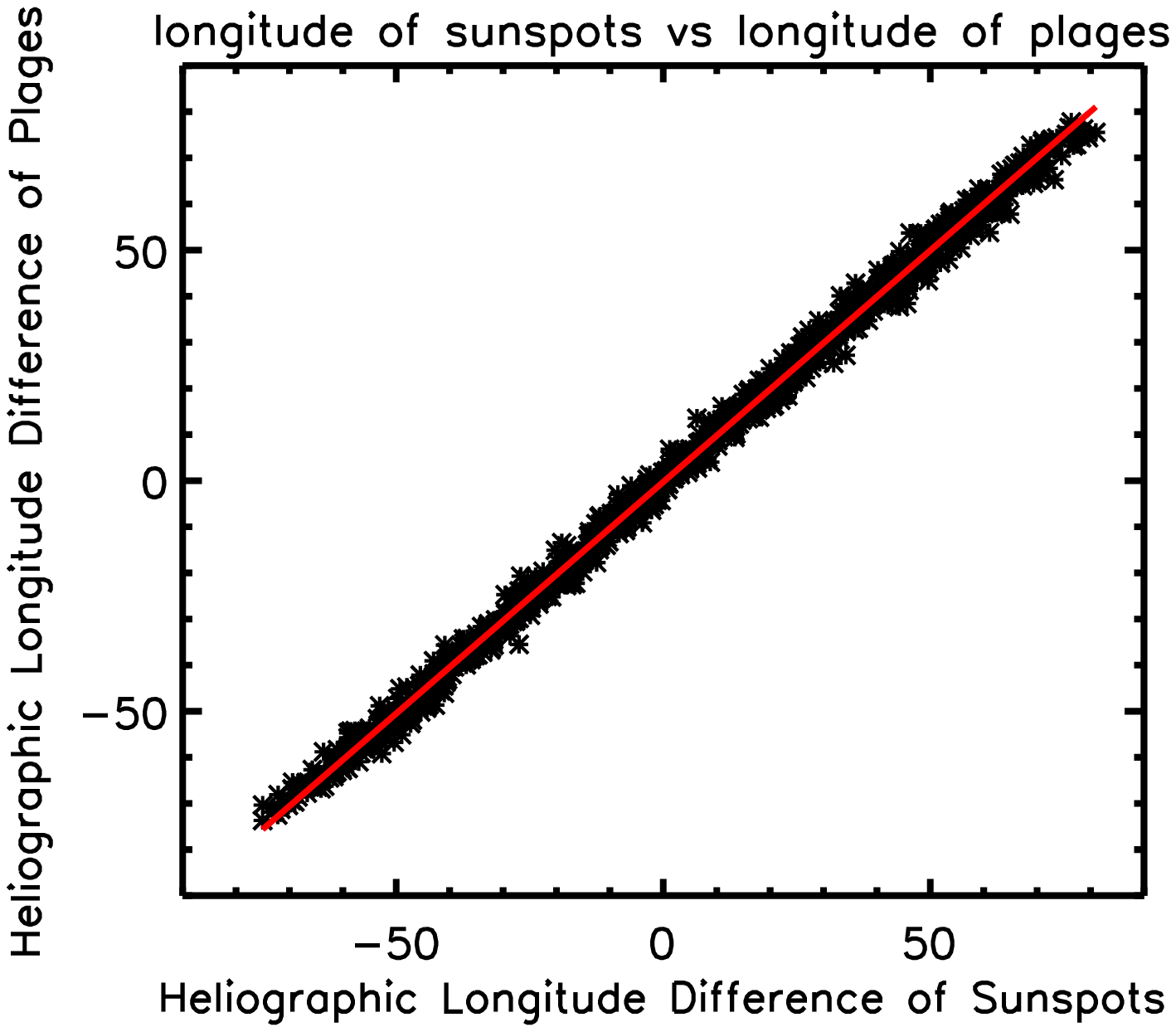}} \\
\end{tabular}
    \caption{ Figure 4a illustrates a scatter plot between estimated average latitude of calcium plage
and sunspot latitude. Figure 4b illustrates a scatter plot between average longitude from the
central meridian of a detected plage and the sunspot.}
\end{center}
\end{figure}

\begin{figure}
\begin{center}
     Figure 5a \hskip 30ex  Figure 5b
%\vskip -6.5ex
    \begin{tabular}{cc}
      {\includegraphics[width=13pc,height=12pc]{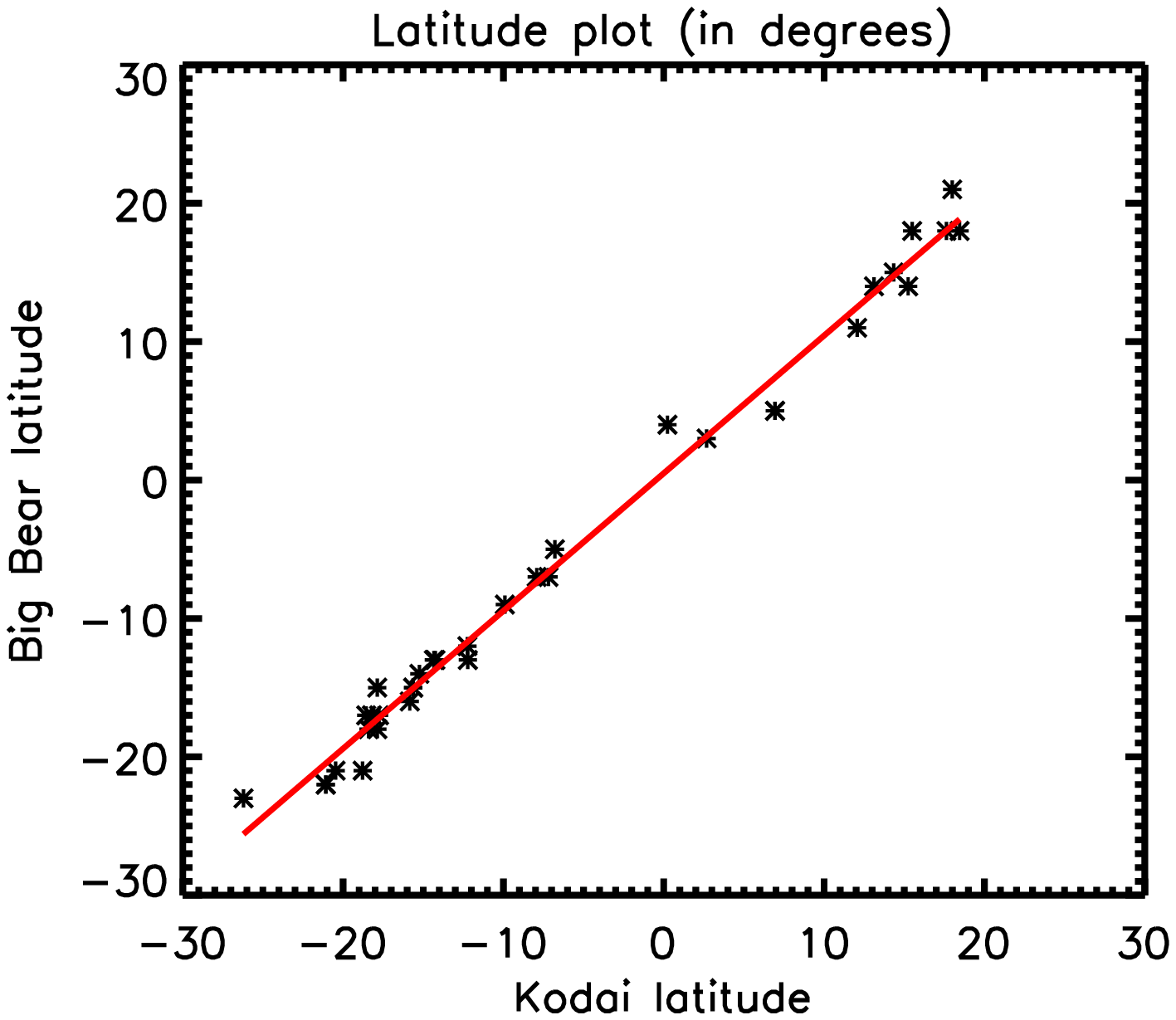}} &
      %\vector(10,10){15}
      {\includegraphics[width=13pc,height=12pc]{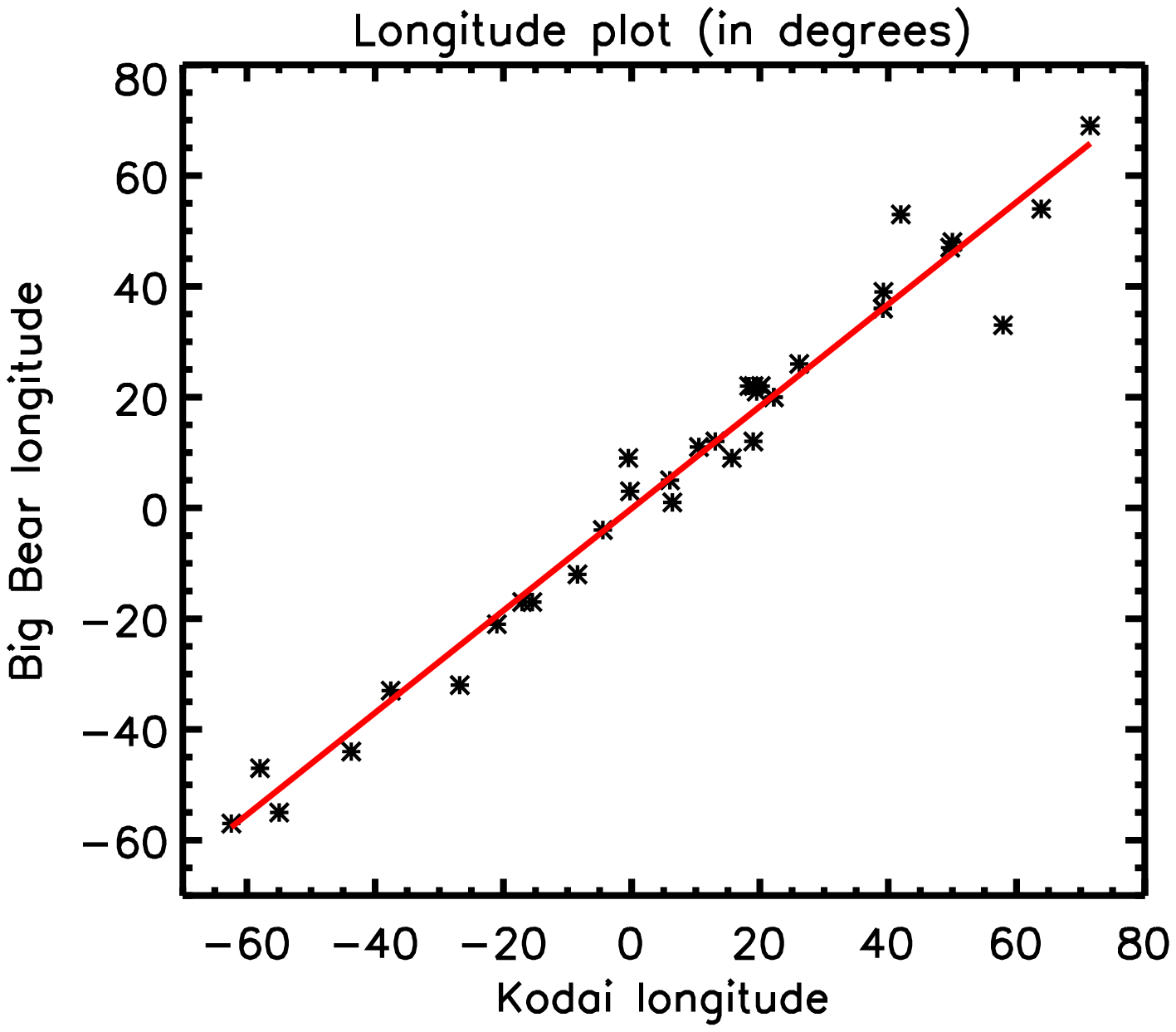}} \\
\end{tabular}
    \caption{ Figure 5a illustrates a scatter plot between estimated Kodai average latitude of calcium plage
with the estimated Big Bear calcium plage latitude. Whereas  5(b) illustrates a scatter plot between
estimated Kodai average longitude of calcium plage with the estimated Big Bear calcium plage longitude.}
\end{center}
\end{figure}

\begin{figure}
\begin{center}
 \begin{tabular}{cc}
      {\includegraphics[width=16pc,height=15pc]{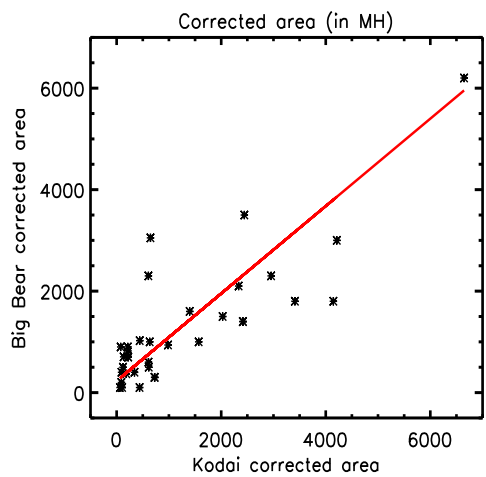}}
\end{tabular}
    \caption{ Illustrates a scatter plot between estimated average area (in mh) of Kodai calcium plage
with the estimated average area (in mh) of Big Bear calcium plage.}
\end{center}
\end{figure}

\begin{figure}
\begin{center}
     Figure 7a \hskip 30ex  Figure 7b
%\vskip -6.5ex
    \begin{tabular}{cc}
      {\includegraphics[width=13pc,height=12pc]{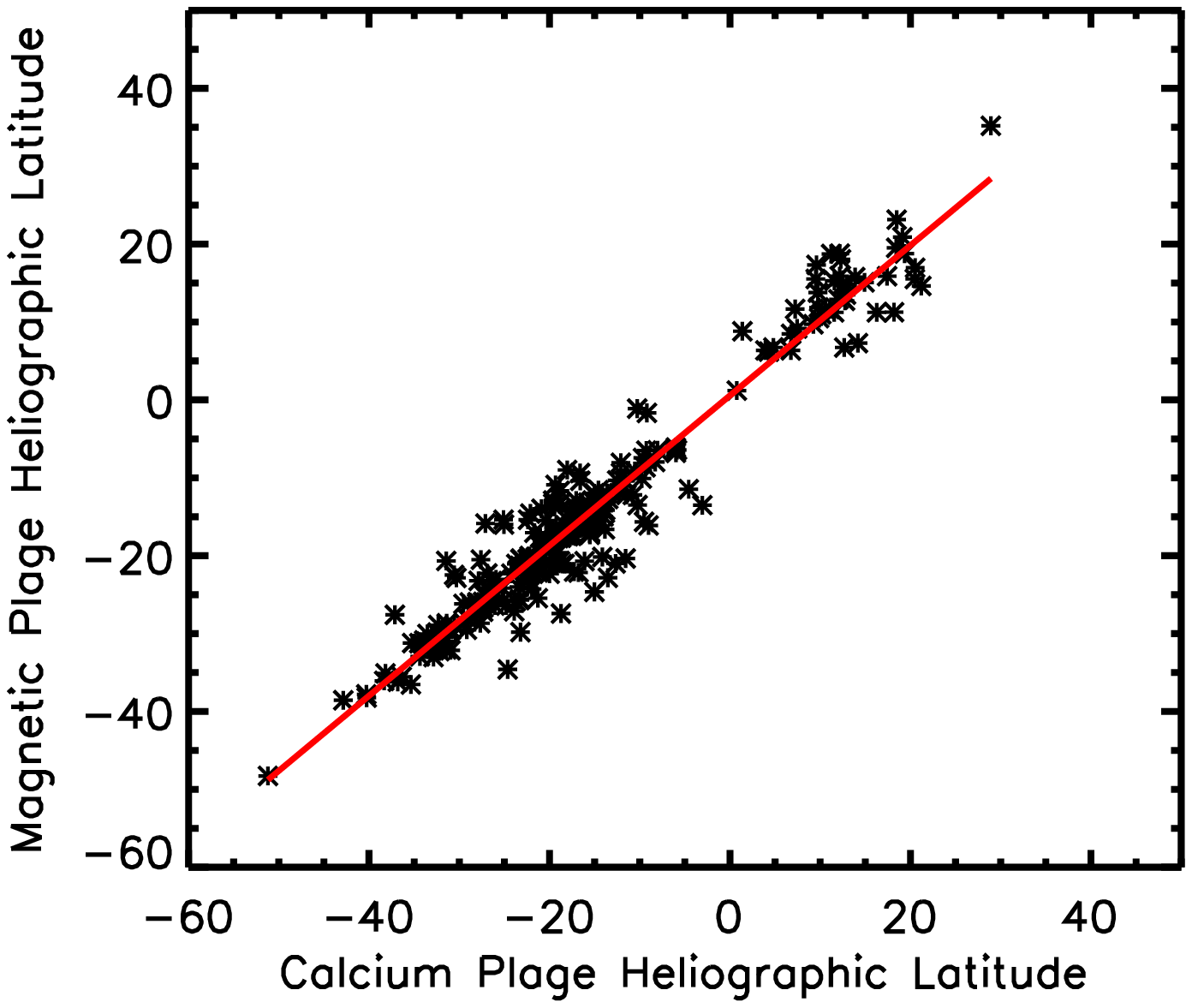}} &
      {\includegraphics[width=13pc,height=12pc]{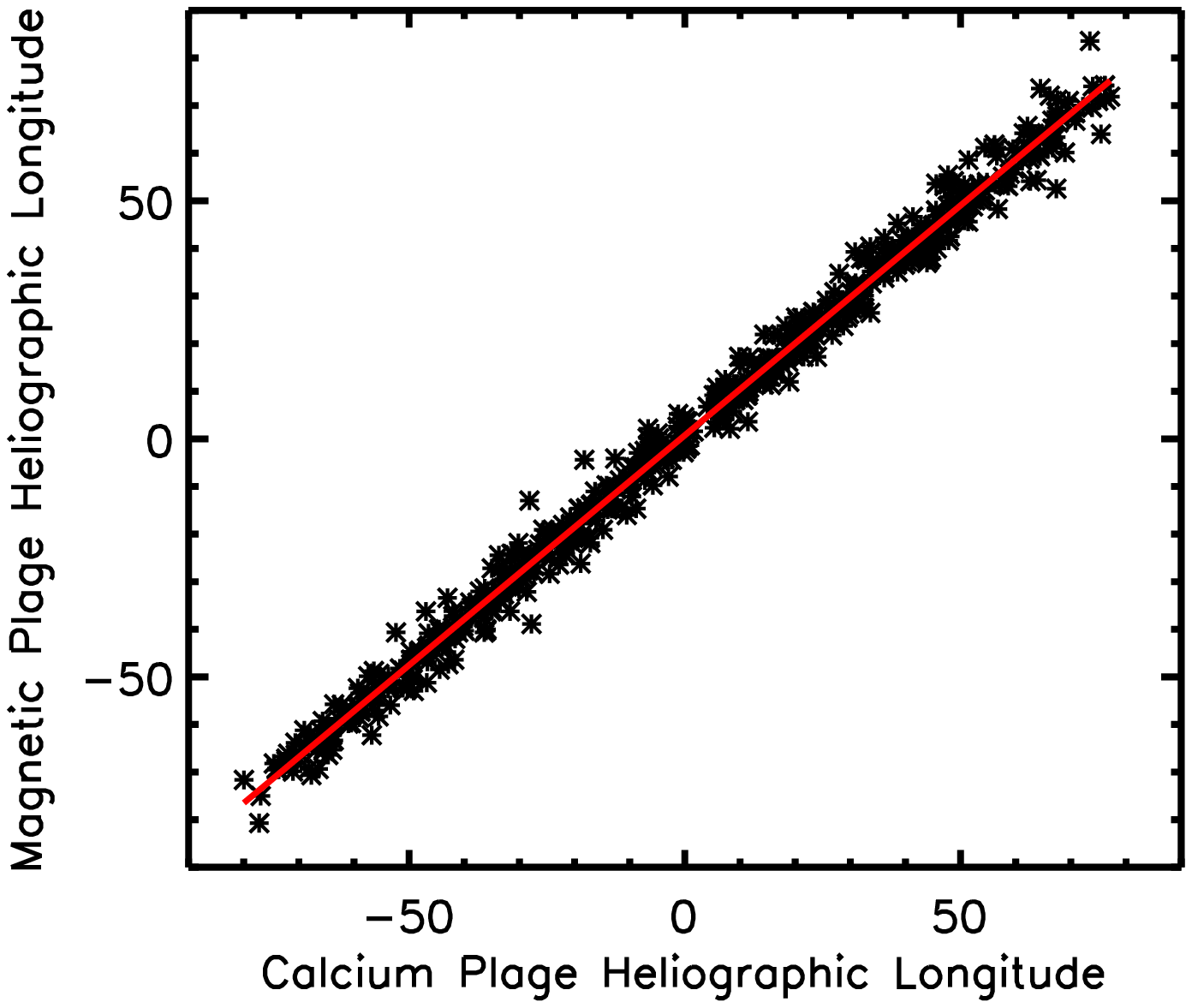}} \\
\end{tabular}
    \caption{ Figure 7a illustrates a scatter plot between estimated average latitude of calcium plage
and magnetic plage (detected from the SOHO/MDI magnetograms) latitude. Figure 7b illustrates a scatter plot between average longitude from the
central meridian of a detected calcium plage and the magnetic plage.}
\end{center}
\end{figure}

\begin{figure}
\begin{center}
%     Fig 4(a) \hskip 40ex  Fig 4(b)
%\vskip -6.5ex
    \begin{tabular}{cc}
      {\includegraphics[width=13pc,height=12pc]{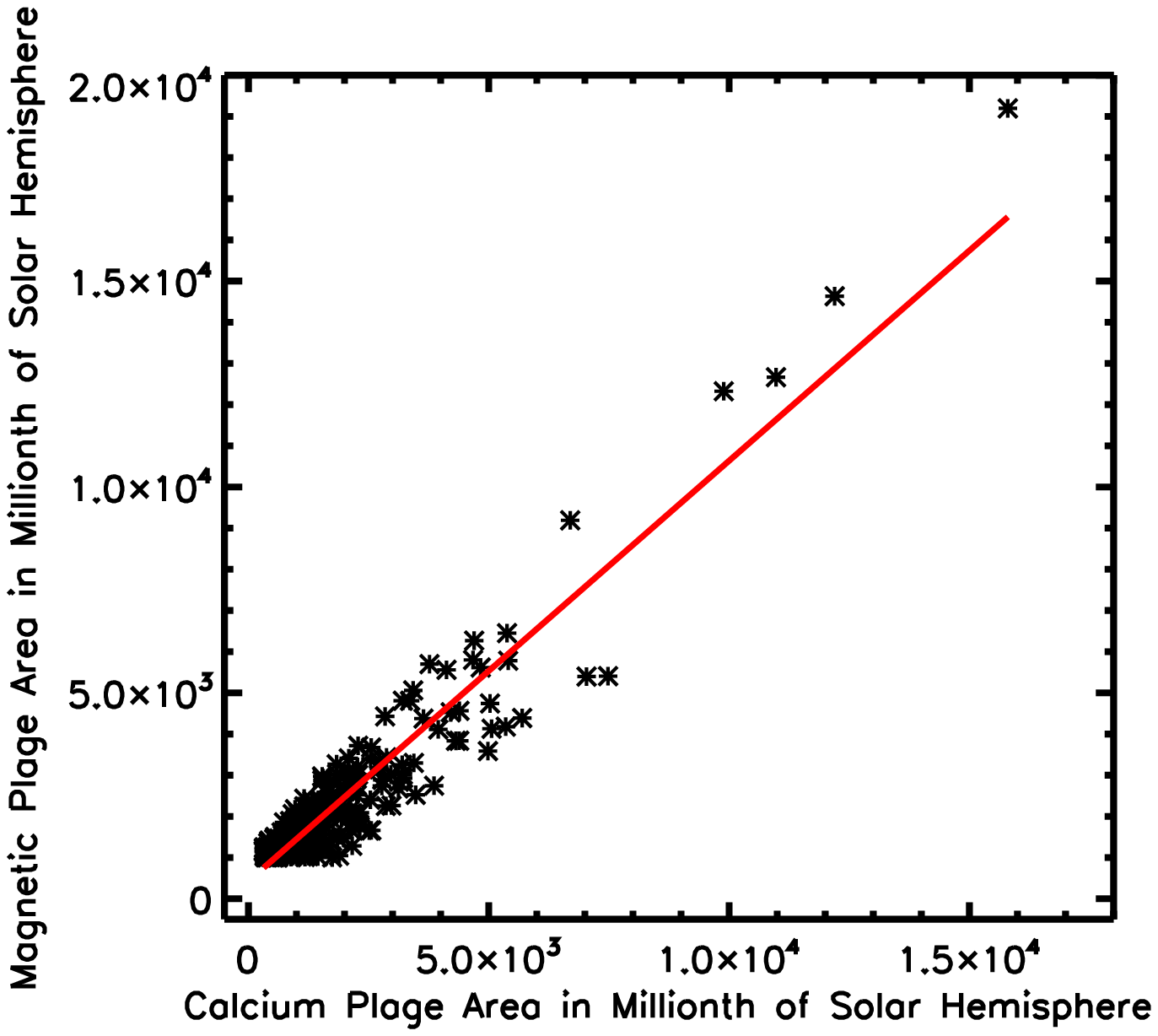}} &
      {\includegraphics[width=13pc,height=12pc]{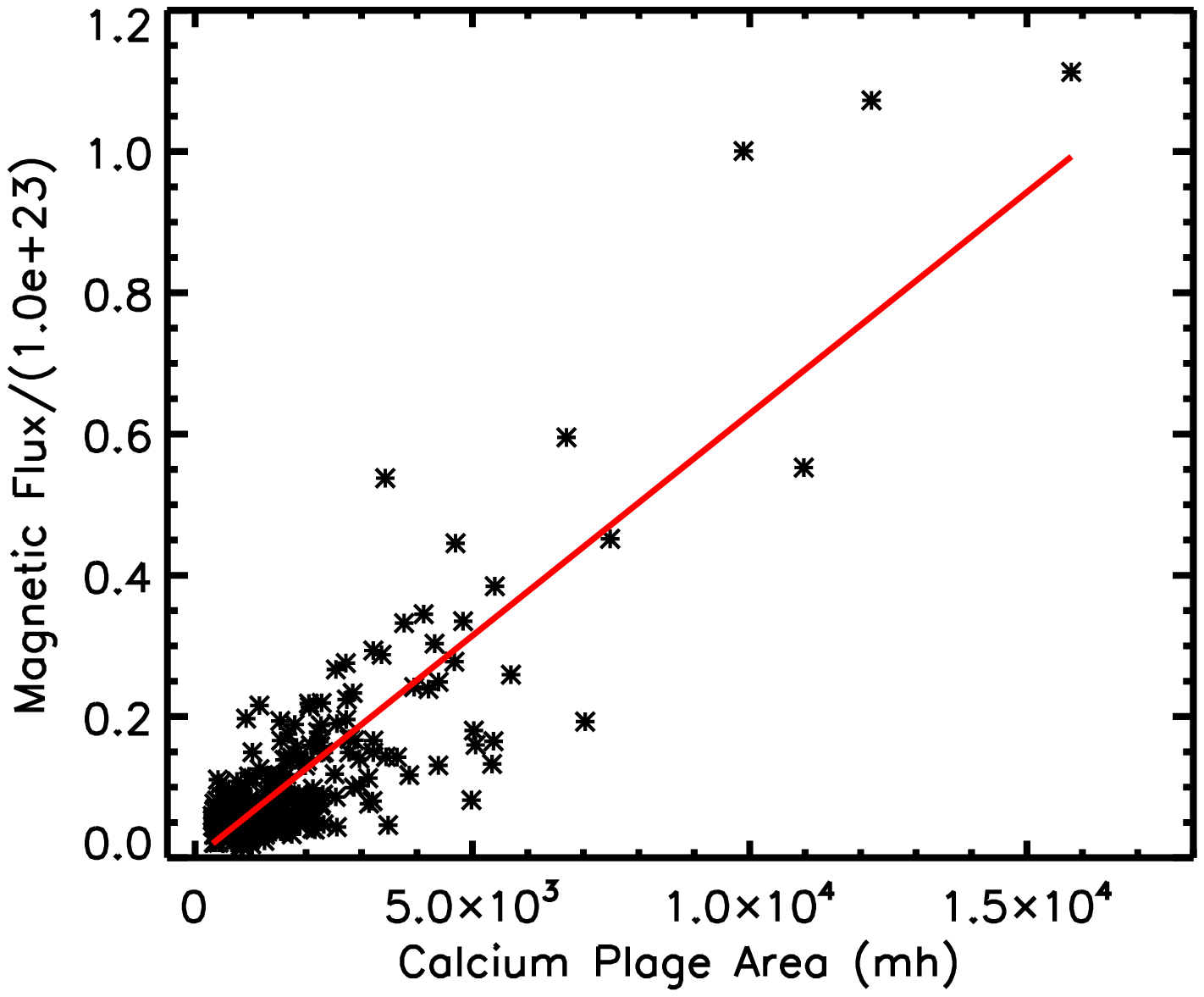}} \\
\end{tabular}
    \caption{ Figure 8a illustrates a scatter plot between estimated average area (in mh) of calcium plage
and magnetic plage area. Figure 8b illustrates a scatter plot between average area (in mh) of calcium plage with
the magnetic flux (normalized with $10^{23}$ Mx) of the magnetic plage obtained by SOHO/MDI magnetograms.}
\end{center}
\end{figure}

\subsubsection {Ellipse Fitting}

Circle fitting can not be unique and validated unless it should be shown that ellipcity
of the image due to atmospheric distortion is negligible.
Since we have the detected edge of the solar disk, we can also calculate the ellipcity of the
solar disk. For this purpose we employ Halir and Flusser (1998) algorithm which modifies Fitzgibbon and Fisher (1999) algorithm for ellipse fitting.
If $a$ and $b$ are equatorial and polar axes, then  ellipcity parameter is defined as $|{(b-a) \over b}|$.
For the Kodai images, estimated ellipcity parameter for annual and monthly data is presented
in Figure 2. One can notice that, as ellipcity ($\sim 0.1 \%$ for the yearly average and $\sim 0.1 \%$ for
the monthly average) of the images is negligible, we are very
much safe in fitting the circle to the observed images.

\subsubsection {Fixing Heliographic Coordinates}

Accurate calculation of heliographic coordinates is necessary, as without
accurate computation of the same it is impossible to do any meaningful science with calcium images.
 Before determining the weighted average heliographic coordinates of detected plages, heliographic
coordinates need to be calculated and assigned for all the pixels of the image.
Following Smith and Zwart (1990), we compute the heliographic coordinates of each pixel with in the detected calcium image as follows.
 With daily known values of heliographic latitude ($B_{0}$)
and longitude ($L_{0}$) of the disk center as well as the polar angle $P$, heliographic longitude
$L$ and longitude difference from central meridian $l$ of each pixel are computed.

If $T = (JD-2415020)/36525$ ($JD$ is the Julian Date of observation and $T$ is the number of Julian
centuries since epoch 1900 Jan 0.5), then geometric mean latitude $L^{'}$, mean anomaly $g$ and right ascension $\Omega$
 of the ascending node of the Sun are

\begin{equation}
L^{'} = 279.69668^{o} + 36000.76892^{o}T + 0.0003025^{o}T^{2} ,
\end{equation}
\begin{equation}
g = 358.47583^{o} + 35999.04975^{o}T - 0.00015^{o}T^{2} - 0.0000033^{o}T^{3} ,
\end{equation}
and
\begin{equation}
\Omega = 259.18^{o} - 1934.142^{o}T .
\end{equation}

\noindent Summation of mean latitude $L^{'}$ and $C$ yields the true longitude $\lambda_{\odot}$ of the Sun
\begin{equation}
\lambda_{\odot} = L^{'} + C,
\end{equation}
here $C$ is equation of the center and is defined as follows
\begin{equation}
C=C_{1}+C_{2}+C_{3} \, ,
\end{equation}
where
\begin{equation}
C_{1}=({1.91946}^{o} - 0.004789^{o}T - 0.000014^{o}T^{2} )sin(g),
\end{equation}
\begin{equation}
C_{2}=(0.020094^{o} - 0.0001^{o}T )sin(2g), 
\end{equation}

\noindent and $C_{3}=0.000293^{o}sin(3g)$ respectively.

\noindent With true longitude $\lambda_{\odot}$ and corrections for aberration and nutation, apparent longitude $\lambda_{a}$ of the
sun is computed as follows
\begin{equation}
\lambda_{a} = \lambda_{\odot} - 0.00569^{o} - 0.00479^{o}sin(\Omega) .
\end{equation}
\noindent Further parameter $\phi$, as given in the following equation, is required for the computation of actual physical ephemeris
\begin{equation}
\phi= {360 \over{25.38}}(JD - 2398220).
\end{equation}
\noindent With inclination ($I= 7.25^{o}$) of sun's axis relative to the ecliptic plane, the
longitude of the ascending node of the solar equator , $K$, is given as follows
\begin{equation}
K = 74.3646^{o} + 1.395833^{o}T .
\end{equation}
where $X$ and $Y$ are
\begin{equation}
tanX = - cos(\lambda_{a} )tan(\epsilon) ,
\end{equation}
and
\begin{equation}
tanY = - cos(\lambda_{\odot} - K )tan(I) ,
\end{equation}
here $\lambda_{a}$ is the sun's apparent longitude corrected for
nutation and $\epsilon$ is the obliquity of the ecliptic.
Then mean obliquity $\epsilon_{0}$ is computed as follows
\begin{equation}
\epsilon_{0} = 23.452295^{o} - 0.0130125^{o}T - 0.00000164^{o}T^{2} + 0.000000593^{o}T^{3} \, .
\end{equation}
With a correction for nutation, mean obliquity $\epsilon$ is given as follows
\begin{equation}
\epsilon = \epsilon_{0} + 0°.00256cos(\Omega) .
\end{equation}
Ultimately daily central coordinates $B_{0}$ and $L_{0}$ and the polar angle $P$ of the sun are computed as follows

\begin{equation}
P =X +Y ,
\end{equation}

\begin{equation}
B_{0} = sin^{-1} [sin(\lambda_{\odot} - K )sin(I)] ,
\end{equation}

\begin{equation}
L_{0} = tan^{-1}[{sin(K-\lambda_{\odot} )cos(I) \over{ -cos(K-\lambda_{\odot} )}}]+M,
\end{equation}
\noindent where $M = 360- \phi$. In this case $\phi$ varies from $0^{o} - 360^{o}$ which can be obtained
by subtracting integral multiples of $360^{o}$ .

Owing to daily orbital changes of the Earth around the sun, resolution of pixels also changes.
Hence, due to this daily orbital variations, distance $R^{'}$ (in AU) of the Earth from the sun
also changes and is given as follows

\begin{equation}
R^{'} = 1.00014 - 0.01671cos(g) - 0.00014cos(2g) .
\end{equation}
The semi-diameter $Rad$ (in arc-seconds) is given as follows
\begin{equation}
Rad = ( {0.2666 \over{R^{'}}})^{o} \times 3600^{''} .
%Rad = ( 0.2666 \over{R^{'}})^{o} \times 3600^{"} .
\end{equation}

Finally computation of heliographic coordinates is achieved through transformation from sun's image
in Cartesian coordinates to the polar coordinates $(r, \theta^{'} )$.
Mathematical determination of the heliographic coordinates is based on the polar
coordinates $(r, \theta^{'} )$. For this purpose one has to compute the
angular distance $\rho$ of any pixel from the center of the solar disk in the following way
\begin{equation}
sin \rho = {r\over{R}} ,
\end{equation}
where $R$ is the radius of the solar disc, as described above in section 2.1.2, estimated by the circle fitting.
Next step is to compute heliographic latitude $\theta$ and longitude $l$ from the central meridian of any pixel as follows
\begin{equation}
sin \theta = cos \rho sin B_{0}  + sin \rho cos B_{0} sin \theta^{'} ,
\end{equation}
and
\begin{equation}
 sin l = {{cos\theta^{'}sin{\rho}}\over cos\theta}.  
\end{equation}

In order to have accurate estimation of heliographic
coordinates, correction for distortion of the projected
image due to the telescope also has to be taken into account. For this purpose projected image can be corrected in the following steps
\begin{equation}
T=Rad/15
\end{equation}
\begin{equation}
R_{0} = {Rad \over{(7 \times 36)}} \times 10^{-12} \times 29.5953cos[ {cos^{-1}(-0.00629T) \over 3}+240],
\end{equation}
\begin{equation}
 \rho^{'} = R_{0} \times {r\over{R}},
\end{equation}

\noindent and
\begin{equation}
\rho= sin^{-1} ({ sin(\rho^{'})\over{sin(R_{0}}))}) - \rho^{'}.
\end{equation}
This $\rho$ is then taken as the corrected angular distance and then the heliographic coordinates
are computed as mentioned above.

Final step is to estimate the heliographic longitude $L$ by adding longitudinal difference $l$ of any pixel from the central meridian to the value of $L_{0}$ such that
\begin{equation}
L =L_{0}+ l.
\end{equation}
\begin{table*}
%\centering
%\begin{minipage}{320mm}
\begin{center}
\caption{ Heliographic coordinates of typical detected calcium plages from the
 Kodai image}
\hskip 8ex
% \begin{tabular}{@{}|lll| ll| ll| @{}}
 \begin{tabular}{@{}lccccc}
 \hline
%\multicolumn{3}{|c|}{CH1} & \multicolumn{2}{|c|}{CH2} & \multicolumn{2}{|c|}{CH3}\\ \hline
%                    CH1 &     CH2   & CH3                                 

Date of     & Lat & Err  & Long & Err   & Corrected \\
Observation  &  (Deg)  &  (Deg)   & (Deg)  & (Deg) & Area (mh) \\

 \hline
1990 1 1.12 & -28.65 & 0.15 & -52.87  & 0.32 &  1060.93 \\
1990 1 1.12 & -28.01 & 0.02 & 58.95 & 0.034 & 5706.50 \\
1990 1 1.12 & -19.92 &  0.02 & 34.03 &  0.02 & 588.70 \\
1990 1 1.12 & -19.56 & 0.01 & 18.73 & 0.04 & 3298.71 \\
1990 1 1.12 & -11.73 & 0.03 & 59.85 & 0.04 & 300.57 \\
1990 1 1.12 & -14.41 & 0.02 & 49.55 & 0.025 & 2289.96 \\
1990 1 1.12 & -11.58 & 0.03 & 36.25 & 0.02 & 2355.40 \\
1990 1 1.12 &  11.53 & 0.02 & -28.93 &  0.03 & 601.20 \\
1990 1 1.12 &  12.44 & 0.01 & 4.87 & 0.05 & 336.26 \\
1990 1 1.12 & 14.10 & 0.02 & -45.80 & 0.03 & 1727.19 \\
1990 1 1.12 & 14.59 & 0.02 & 14.63 & 0.04 & 419.23 \\
1990 1 1.12 & 24.60 & 0.02 & 35.43 & 0.04 & 1413.40 \\ \hline

\end{tabular}
%\end{minipage}
\end{center}
\end{table*}

\subsubsection{Morphological Operations For Plage Detection}

Morphological operations are required for detection of plages as they ensure accuracy of the
average heliographic coordinates and the area of detected plages.
Bilateral filtering is performed before a copy of the spectroheliogram is run through OpenCV's
findContours function. This operation is similar to Gaussian filtering such that the image
histogram is smoothed, but in Bilateral filtering edges are not smoothed. This allows for
detected plages to have similar pixel intensities, but ensures that plage intensities are not
redistributed into the background quiescent area of the sun.

Above step is followed by thresholding, which converts the image into a two level colour
image. A value for threshold is provided, and all pixel values above this value are converted
to 255 (black) while all pixel values below the threshold are converted to 0 (white). The
threshold itself is calculated by calculating the median and standard deviation of all pixels
lying inside the solar disk for each image. The final threshold is set to the median plus 3
standard deviations. We also tried with a threshold of mean plus 3 standard deviation. In the
following we find that both the thresholds yield same results. In Figure 3a we illustrate
the original calcium image and in Figure 3b detected plages in the same image is illustrated.

Other morphological operations such as opening (erosion followed by dilation) were also
considered, but were avoided since they often split one detected plage into two detected
plages. This happened in particular for large plages connected by thin tendrils.

\subsubsection{ Computation Of Average Heliographic Coordinates}

Following Hiremath and Hegde (2013), weighted positional coordinates (heliographic latitude $\theta_{plage}$ and the longitude from the
central meridian  $l_{plage}$) with their error bars ($\delta \theta$ and $\delta l$) are computed as follows.
If intensity $I_{n}$ is n-th pixel of the detected plage, positional coordinates of the plage are the weighted
averages which are estimated in the following way
\begin{equation}
\theta_{plage} = {\sum_{n} \theta_{n} I_{n} \over{ \sum_{n} I_{n}}}
\end{equation}
and
\begin{equation}
l_{plage} = {\sum_{n}{ l_{n} I_{n}} \over{ \sum_{n}{I_{n}}}}
\end{equation}

\noindent Similarly errors $\delta \theta$ and $\delta l$ in the in the positional coordinates are obtained
as follows
\begin{equation}
\delta \theta = {\sigma_{\theta} \over{ \sqrt(N)}} ,
\end{equation}
and
\begin{equation}
\delta l = {\sigma_{l} \over{ \sqrt (N)}}.
\end{equation}

Where $\sigma_{\theta}$ and $\sigma_{l} $ represent, with in each detected plage, standard deviations of latitude and longitude from the central
meridian and $N$ is the total number of pixels in the detected plage.
Using these formulae, the heliographic coordinates and their errors for each plage in every
image are estimated.

For the authentication of our detected plages and their estimation of average positional coordinates,
we assume that well developed
calcium plages are extension of sunspot flux tubes in the chromosphere. Plage heliographic
coordinates are then compared with the heliographic coordinates of the sunspots (in this
case we consider Greenwich sunspot heliographic coordinates data) at the
photospheric level. In case the observed timings of sunspots are different compared
to the observed timings of the Kodai detected plages, we also account
for change in the longitudinal difference of sunspots due to rotation of the sun as follows
\begin{equation}
l = l_{0} + \Omega \delta t ,
\end{equation}
and
\begin{equation}
\Omega = \Omega_{0} - \Omega_{1} sin^{2} \theta,
\end{equation}
where $\delta t$ is the difference of the time of observations, $l$ is the weighted average longitude
difference from central meridian, corrected for difference in observation time of sunspot and
plage, $l_{0}$ is the calculated weighted average longitude difference, $\Omega$ is the rotation rate of the
sun and, $\theta$ is the weighted average heliographic latitude for the plage. In the above
equation, $\Omega_{0}$ and $\Omega_{1}$ are the constant coefficients that are due
to equatorial and high latitude rotation rate of the sunspots. Essentially,
this formula accounts for differential rotation rate of the sun.

\subsubsection{Projected And Corrected Plage Areas}

For area estimation, we use the OpenCV function contourArea, which determines the area of
each contour in pixels. Since our detected contours are plages, this represents plage area in
pixels. By knowing resolution (incase of Kodai Ca II image it is 0.85") of the image, area of
each pixel is computed and total area of plage is a sum of area of all the pixels within the
detected plage boundary. This area is converted to millionths of solar hemisphere area using
the uniquely estimated solar radius. As the sun is a sphere, plages that occur
near the solar edges and high latitudes require correction for the projectional
effects which is given as follows
\begin{equation}
A = {A^{'} \over cos\delta} \, , 
\end{equation}

\noindent where $A^{'}$ is projected area and $cos \delta = sin B_{0} sin \theta + cos B_{0} cos \theta cos l$,
with estimated average positional coordinates ($\theta$ and $l$) and $B_{0}$ is the heliographic latitude
of center of the solar disk at the time of observation.

\section{Results}
In order to authenticate the estimated positional coordinates and the areas of detected calcium plages, we have
following three comparisons: i) photospheric sunspots positional coordinates
with the positional coordinates of the chromospheric calcium plages,
ii) Kodai calcium positional coordinates and area with the Big-Bear calcium
positional coordinates and the area and,
iii) positional coordinates and area of the chromospheric magnetic plages (extracted from SOHO/MDI magnetograms)
with the positional coordinates and area of Kodai calcium plages,

\subsection{Comparison Of Positional Coordinates Of Chromospheric Plages With Sunspot Positional Coordinates}

As there is a one-to-one correspondence (Ortiz and Rast 2005, Sivaraman and Livingston 1982) between calcium plages and sunspots,
we compare the calculated average positional coordinates ($\theta$ and $l$) of plages detected between 1909 and 2007 with Greenwich sunspot
data (https://solarscience.msfc.nasa.gov/ \, \, greenwch.shtml).
We find that for observations taken on
the same date the heliographic latitudes match very well within a limit of one degree. The
calculated heliographic longitude difference from central meridian also matches with the
longitude difference of the sunspots within a limit of one degree.
 These results are illustrated in Figures 4a and 4b respectively.
Best way of comparison and authentication is to compare the heliographic coordinates and the area
of Kodai calcium plages with the heliographic coordinates and area of calcium plages estimated
from other studies. For this purpose, Big Bear Observatory calcium plages are compared (Figures 5 and 6 ).
We find that estimated heliographic coordinates and corrected area (in millionth of hemisphere) of Kodai
calcium plages match very well with the Big Bear Observatory estimated heliographic coordinates
and corrected area respectively that clearly authenticates our method of detection and estimation
of position and area of calcium plages.

For the year 1990, in the month of Jan, estimated average heliographic coordinates and the
 corrected area of typical plages are presented in Table 1. The first column represents, year, month and
date of observation (in decimal). Second and third columns are the estimated latitude and its error.
Similarly fourth and fifth columns are estimated longitude from the central meridian and its error.
Whereas the last column represents the corrected area of the plage.

\subsection{Comparison Of Positional Coordinates And Area Of Calcium Plages With Positional Coordinates
 And Area Of Chromospheric Magnetic Plages}
It is well established fact that, compared to sunspots, plages decay slowly. Hence, there
is every possibility that at least in some cases, sunspots positions may not match very well with the
calcium plages. Hence, best way is to compare positional coordinates and the area of calcium plages
with the  positional coordinates and the area of chromospheric magnetic plages
as these activity regions originate near the same region of chromosphere.
For such comparison, SOHO/MDI magnetograms are ideally best suited. This is due to the fact that
the observed Ni I 6768 $\AA$ line (Scherrer {\em et.al.} 1995) used for generating the SOHO/MDI magnetograms
occurs (Jones 1989, Meunier 1999) at about 200 Kms above the photosphere, very near the Ca II line
height formation (see Figure 2 of Yang {\em et.al.} 2009 and, Table 1 of  Anusha and Nagendra (2013)). In order to
achieve this  aim, we use 96 min averaged SOHO magnetograms, a similar method (as described in the above section 2)
is used to detect magnetic plages, fix the heliographic coordinates and estimate the area.
By using magnetic field intensity of 30 G (two times noise
level in the magnetogram; this threshold is to detect correct boundary of the magnetic plage and
not the extended diffuse structure) as threshold, we detect magnetic plages.
%For comparison, both
%the detected magnetic plages (derived from magnetogram) and Kodai calcium plages are
%presented as.... 
Heliographic coordinates and areas of detected magnetic and calcium plages
are illustrated in Figures 7 and 8a respectively. As for the area of detected
 magnetic plages, although there are positive and negative magnetic flux values, negative
flux value pixels are converted into absolute (positive) values and their areas
are added to the areas of the positive flux pixels.
 It is interesting to note that
 heliographic coordinates and area of calcium plages
match very well with the heliographic coordinates and area of magnetic plages
authenticating our method of detection and estimation of position and
areas of calcium plages.

As described in the introduction, our one of main aims is to investigate long-term ($\sim$ 100 yrs) variation
of solar magnetic activity. From this study we find that, as there is one to one correspondence between the position and heliographic
coordinates, calcium plage area can be used as a magnetic proxy. That means if one finds a relationship between
area of calcium plage and magnetic flux of the magnetic plages, nearly 100 years of Kodai calcium area data
can be used to investigate the long-term variation of solar magnetic activity. For this
purpose, within a detected magnetic plage boundary, negative magnetic flux are converted into absolute (positive) flux values
and are added with rest of the positive flux values in order to estimate the total magnetic flux. A relationship
between calcium plage area and magnetic plage total flux is illustrated as a scatter plot in Figure 8b.

\section{ Conclusions}
A code is developed in Python to analyze Kodaikanal Ca II K spectroheliograms.
For determination of center and radius of solar disk uniquely, a circle is
least square fitted to the detected edge of the solar disk. Further, we find that the Kodaikanal Ca II K Spectroheliograms have a
mean ellipcity of 0.0010, which suggests application of circle fitting is reasonable for the
Kodai Ca II K spectroheliograms. By using standard astronomical ephimerides, for each
pixel of the solar disc, heliographic coordinates such as latitude and longitude are computed.
By applying image processing techniques, plages of the chromosphere are detected. This
process is optimized using parallel programming. Weighted average heliographic
coordinates and area of the plages are computed. From two methods,  heliographic
coordinates and area of the calcium plages are compared with heliographic coordinates
of sunspots and magnetic plages that shows a one to one correspondence and validating
our method of detection of Kodai calcium plages.

\centerline{\bf Acknowledgments}

SOHO is a project of international cooperation between ESA and NASA.

\vskip 0.5cm

\centerline{\bf REFERENCES}
\vskip 0.5cm
 Alfv{\'e}n, H. 1943, Arkiv f. Mat., Astron. o. Fys., 29A (12), p. 1-17

 Anusha, L.~S. and Nagendra, K. N. 2013, ApJ, 767, 108

 Balmaceda, L. A, Solanki, S. K, Krivova, N. A, and Foster, S. 2009, Journal of
   Geophysical Research: Space Physics 114, A7

 Barata, T., Carvalho, S., Dorotovič, I.,  Pinheiro, F. J. G., Garcia, A., Fernandes, J \& Lourenço, A. M., 
     Astronomy and Computing, 2018, 24, 70

 Bertello, L., Ulrich, R. K., and Boyden, J. E.: 2010, Solar phys, 264, 31

 Bertello, L.; Pevtsov, A. A.; Tlatov, A. G., in Proceedings of a Meeting held at the University of Coimbra, 
   Edited by Ivan Dorotovic, Catherine E. Fischer, and Manuela Temmer. ASP Conference Series, Vol. 504. 2016a, p.213

 Bertello, L., Pevtsov, A. A., Tlatov, A \& Singh, J., 2016b, Solar phys, 291, 2967 

 Butler, C. J., in IAU 176, 1995, Edited by Klaus G. Strassmeier and Jeffrey L. Linsky, Kluwer Academic Publishers, Dordrecht, p.423

 Chatterjee, S, Banerjee, D and Ravindra, B. 2016, ApJ, 7, 87

 Chatzistergos, T., 2017, Ph.D thesis, University of Gottingen

 Chatzistergos, Theodosios, Ermolli, Ilaria; Solanki, Sami K.; Krivova, Natalie A., 2018, Astron Astrophys, 609, 92

 Ermolli, I, Marchei, E, Centrone, M, Criscuoli, S, Giorgi, F and Perna, C.: 2009a, Astron Astrophys, 499, 627

 Ermolli, I, Solanki, S. K, Tlatov, A. G, Krivova, N. A, Ulrich, R. K and Singh, J.
  2009b, ApJ, 698, 1000

 Foukal, P., 1996, GRL, 23, 2169

 Frasca, A., Freire Ferrero, R.,  Marilli, E., Catalano, S, et.al., 2000, Astron Astrophys, 364, 179 

 Frasca, A.,  K. Biazzo, K., Taş, G., Evren, S and A. C. Lanzafame, A. C, 2008, Astron Astrophys. 479, 557

Frasca, A.,  Biazzo, K., . Kővári, E. Marilli and Ö. Çakırlı, 2010, Astron Astrophys, 518, A48
 A. Fitzgibbon, M. Pilu, and R. B. Fisher. Direct least square fitting
of ellipses. 1999, IEEE Transactions on Pattern Analysis and Machine Intelligence, 21(5), 476

 R. Halir and J. Flusser. Numerically stable direct least squares fitting
of elllipses. In Proc. of Sixth Int’l Conf. 1998, Computer Graphics and
Visualization, 1, 125

 Gokhale, M. H, Javaraiah, J and Hiremath, K. M. 1990, IAU symp, 138, 375

Hiremath, K. M. 1994, Ph.D thesis, Bangalore University, India

Hiremath, K. M. and Gokhale, M.~H. 1995, ApJ, 448, 437

Hiremath, K. M. 2009, Sun and Geosphere, 4, 16

Hiremath, K, M. 2010, Sun and Geosphere, 5, 17

Hiremath, K. M an Hegde, M. 2013, ApJ, 763, 1371

Hiremath, K.~M. 2015, New Astronomy, 35, 8

Jones, H.~P. 1989, Solar phys, 120, 211

Loukitcheva, M., Solanki, S. K.  and White, S. M., 2009, Astron Astrophys, 497, 273

 Meunier, N. 1999, ApJ, 527, 967

 Rodono, M, Byrne, P. B, Neff, J. E, Linsky, J. L, Simon, T, Butler, C. J, Catalano, S, Cutispoto, G, Doyle, J. G,
Andrews, A. D and Gibson, D. M., 1987, Astron Astrophys, 176, 267

 Smith, P. R and Zwart, J. 1990, in Practical Astronomy with your calculator, Cambridge University Pres

 Ortiz, A and Rast, M, 2005, Memorie della Societ Astronomica Italiana, 76, 1018.

 Pevtsov, A.A., Virtanen, I., Mursula, K., Tlatov, A., Bertello, L.: 2016, Astron Astrophys,  585, A40.

 Priyal, M, Singh, J, Ravindra, B, Priya, T. G and Amareswari, K. 2014, Solar phys, 289, 137 

 Priyal, M, Singh, J, Ravindra, B and Rathina, S.~K. 2017, Solar phys, 292, 85

 Scherrer, P. H, Bogart, R. S, Bush, R. I,
       Hoeksema, J. T, Kosovichev, A. G, Schou, J, Rosenberg, W, Springer, L, Tarbell, T, D,
        Title, A. Wolfson, C. J, Zayer, I and MDI Engineering Team, 1995,  Solar phys, 162, 129

 Sivaraman, K.~R. and Livingston, W.~C. 1982, Solar phys, 80, 227

 Tlatov, A. G, Pevtsov, A. A and Singh, J. 2009, Solar phys, 255, 239

 Yang, S.~H. and Zhang, J. and Jin, C.~L. and Li, L.~P. and Duan, H.~Y. 2009, Astron Astrophys, 501, 745

\end{document}